# Hybrid nanocomposites with tunable alignment of the magnetic nanorod filler

*Fabien Perineau[1,2,3], Céline Rosticher[1,2,3], Laurence Rozes[1,2,3], Corinne Chanéac[1,2,3*], Clément Sanchez[1,2,3], Doru Constantin[4], Ivan Dozov[4], Patrick Davidson[4*] and Cyrille Rochas[5]*

[1] UPMC Univ Paris 06, UMR 7574, Chimie de la Matière Condensée de Paris, Collège de France, 11 place Marcelin Berthelot, 75231 Paris Cedex 05, France

[2] CNRS, UMR 7574, Chimie de la Matière Condensée de Paris, Collège de France, 11 place Marcelin Berthelot, 75231 Paris Cedex 05, France

[3] Collège de France, Chaire de Matériaux Hybrides, UMR 7574, Collège de France, 11 place Marcelin Berthelot, 75231 Paris Cedex 05, France

[4] Laboratoire de Physique des Solides, UMR 8502 CNRS, Université Paris-Sud, Bât. 510, 91405 Orsay Cedex, France

[5] CERMAV, UPR 5301 CNRS, BP 53, 38041 Grenoble, Cedex 9, France

Corresponding authors: patrick.davidson@u-psud.fr (33 1 69 15 53 93), corinne.chaneac@upmc.fr (33 1 44 27 15 29)

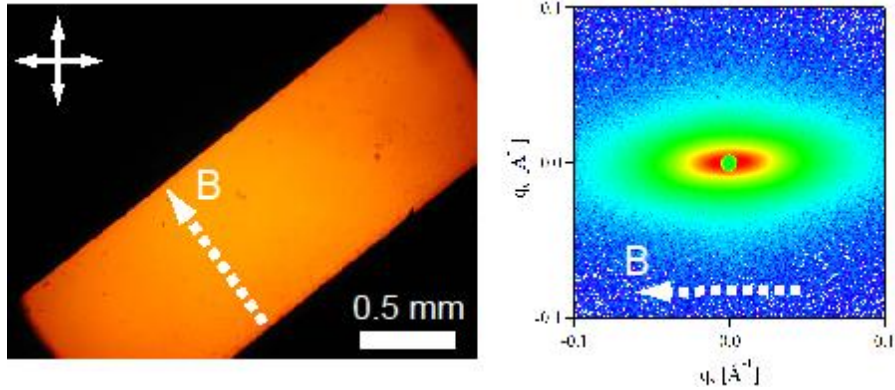

ABSTRACT.

For many important applications, the performance of (polymer – anisotropic particle) nanocomposite materials strongly depends on the orientation of the nanoparticles. Using the very peculiar magnetic properties of goethite ($\alpha$-FeOOH) nanorods, we produced (goethite – PHEMA) nanocomposites in which the alignment direction and the level of orientation of the nanorods could easily be tuned by simply adjusting the intensity of a magnetic field applied during polymerization. Because the particle volume fraction was kept low (1-5.5 vol%), we used the orientational order induced by the field in the isotropic phase rather than the spontaneous orientational order of the nematic phase. At the strongest field values (up to 1.5 Tesla), the particles exhibit almost perfect anti-nematic alignment, as measured by optical birefringence and small-angle X-ray scattering. The results of these two techniques are in remarkably good agreement, validating the use of birefringence measurements for quantifying the degree of orientational order. We also demonstrate that the ordering induced by the field in the isotropic suspension is preserved in the final material after field removal. This work illustrates the interest, for such problems, of considering the field-induced alignment of anisotropic nanoparticles in the isotropic phase, an approach that is effective at low filler content, that avoids the need of controlling the nematic texture, and that allows tuning at will the orientation level of the particles simply by adjusting the field intensity.

1. **Introduction.**

Hybrid organic-inorganic functional materials are currently popular for innovative applications in various domains such as transportation, construction, sealant, and textile.[1-2] The attractiveness of hybrid materials arises from the possibility of combining the properties of each constituent and, in some cases, of even inducing a new property thanks to synergetic effects. In this study, we focus on one class of hybrid materials, namely hybrid Polymer NanoComposites (PNCs), obtained by dispersing inorganic nanoparticles in a polymer matrix. Due to the small particle size, such materials display a huge amount of interfaces provided that a good dispersion is achieved.[3] With nanoparticles, an improvement of the properties of the composite materials is expected for lower filler volume fraction (typically 1-10%) than using traditional micrometer-size filler (15-20%), which is very important since any significant weight reduction would be highly beneficial to the transportation industry for example.[4] Beyond the lower loadings and the quality of nanoparticle dispersion in the polymer matrix, the great challenge in the field of PNCs, in order to improve technological applications, is the ability to control the organization and the distribution of filler to obtain ordered nanocomposites.[5] Indeed, the preferential orientation of anisotropic filler particles is known to induce remarkable optical and mechanical properties.[6-7]

Two main kinds of PNCs, based either on oxide or on clay nanoparticles, are currently being investigated for their specific properties (reinforcement, electric insulation, gas/liquid barrier, flame retardancy).[8-10] More specifically, in a previous article, we reported on the enhancement of the mechanical properties of a PHEMA (poly(hydroxyethyl methacrylate)) matrix reinforced with goethite ($\alpha$-FeOOH) nanorods.[11] By combining different characterization techniques and methodologies, this mechanical reinforcement was attributed to strong interactions, at the iron oxide–PHEMA interface, between the polymer chains and the particle hydroxyl groups, combined with a quite homogeneous dispersion of the nanoparticles.[11] Moreover, nanocomposite spin-coated films, filled with goethite nanorods, displayed birefringence, which was explained by a flow-induced alignment, during the deposition process, of the nanorods within the organic glassy matrix. From an applied perspective, such PNCs are very interesting because the nanofillers are both cheap and not toxic. A remaining challenge in this domain is the precise control of the alignment degree of the nanorods in the composites. Two main approaches can be used to address this issue:[3] (i) the organization of the nanoparticles by applying an external field during matrix polymerization and material processing (called "external-in" approach); (ii) exploiting the spontaneous mesophase formation by anisotropic nanoparticles ("internal-out" approach).[11] The first approach is better suited to low filler contents and is very convenient for goethite-based PNCs because goethite nanorods are easily aligned by magnetic or electric fields, even at low concentrations.[12]

Goethite nanorods, about 400 nm long and 30 nm wide, can easily be synthesized and suspended in water in order to produce stable colloids.[13] Moreover, depending on concentration, these suspensions spontaneously form various liquid-crystalline phases.[14-15] Indeed, beyond some concentration threshold that depends on particle dimensions, goethite nanorods self-organize in a nematic phase.[12, 16] Within a nematic single domain, all rods spontaneously tend to align in the same direction, in contrast with the more dilute, isotropic phase in which the rods point in random directions.

Furthermore, because of their antiferromagnetic structure, goethite nanorods have outstanding magnetic properties. On the one hand, they bear a remanent magnetization due to uncompensated surface spins but, on the other hand, their magnetic susceptibility is weaker along their main axis than in the perpendicular plane.[12] The remanent magnetization gives rise to a linear contribution in the field intensity to the magnetic energy of the nanorods and this contribution is minimum when the rods are parallel to the field. In contrast, the anisotropy of magnetic susceptibility gives rise to a quadratic contribution in the field intensity and that contribution is minimum when the rods are perpendicular to the field. Therefore, in suspension, goethite rods align parallel to a small magnetic field but they reorient perpendicularly at larger field intensities. The crossover between these two behaviors occurs around 0.35 T, independently of the nanorod concentration and organization. At this precise value of the field intensity, the samples are optically isotropic. Consequently, the nematic phase is easily aligned by a small magnetic field and its alignment axis can be switched from parallel to perpendicular by increasing the field intensity.[17]

The magnetic field has also a strong influence on the isotropic phase because of the intrinsic properties of the goethite particles. Indeed, the field induces orientational order in the isotropic phase that becomes birefringent, even at rather small goethite concentrations.[18] Considering the isotropic phase of goethite suspensions instead of the nematic phase allows using smaller filler contents, dispenses with the need of controlling the texture of the samples before applying the field, and provides lower visco-elasticity and response times. In this article, we report on the alignment of goethite nanorods induced within PNCs by a magnetic field, which allowed us to produce hybrid materials whose direction and level of anisotropy is directly controlled by the field intensity.

## 2. Experimental section.

**Inorganic Nanoparticles**. Goethite nanorods were obtained by a dissolution–crystallization process from 2-line ferrihydrite, a poorly-defined highly hydrated phase.[19] 16.6g of $Fe(NO_3)_3, 9H_2O$ were dissolved in 400 ml of distilled water, then the pH of this solution was adjusted to 11 with NaOH (1M) solution. Due to the hydrolysis of iron cations, a brown ferrihydrite precipitate appears. After 20 days of aging, the precipitate became ochre, the characteristic color of goethite nanocrystals. Nanoparticles were washed three times with distilled water, then with a $HNO_3$ (3M) solution to obtain positively-charged particles. Stable aqueous suspensions of non-aggregated goethite particles were obtained by repeated centrifugations of this mixture and dispersions in water. The final pH of the colloidal suspension was close to 3. At pH 11, corresponding to the synthesis conditions, the surface is poorly charged and the particles are flocculated. After centrifugation and peptization of particles in nitric acid, a stable colloidal suspension is obtained. The pH is then close to 3 and the surface charge density of goethite nanorods is +0.2 $C/m^2$.[18] The particle volume fraction (i.e. the ratio of the volume of goethite nanorods to the total sample volume) was $\phi_G$ = 9.4%. Goethite nanoparticles have a lath-like shape with aspect ratio L/D close to 10, with average length of 300 nm, width of 30 nm, and thickness of 10 nm. For these particle dimensions, the onset of the nematic order is around 7% so that all suspensions with nanorod volume fraction smaller than this value are isotropic.

**Polymer matrix.** Poly(2-hydroxyethyl methacrylate) (PHEMA) has been chosen as a glassy polymer matrix for its hydrophilic character, its transparency, and its absence of toxicity. In this study, PHEMA is obtained by UV-light polymerization of 2-hydroxyethyl methacrylate (HEMA, Aldrich, > 99%) monomer added to 2 wt% of a photo-initiator (2,2 Dimethoxy-2-phenylacetophenone (DMPA), Aldrich, 99%).

**Hybrid nanocomposite precursor suspension.** PHEMA-goethite nanocomposites have been prepared as previously described.[11] A goethite aqueous suspension was mixed with various amounts of HEMA (that already contains 2 wt. % DMPA) to obtain several series of samples with volume fraction, $\phi_H$, of goethite particles increasing from 0 to 5.58 %, in hybrid composites. Homogeneous stable hybrid suspensions were thus produced and could readily be polymerized under UV-light to form nanocomposites.

**Preparation of nanocomposite samples.** After vigorous shaking, the suspensions were introduced into flat, 0.05 mm thick, glass capillaries (VitroCom, NJ, USA) that were flame-sealed afterwards. Then, all samples were polymerized under a UV lamp at 25°C for 30 minutes ($\lambda$ = 254 nm, power: 30 W). A magnetic field, ranging from 0 T to 1.5 T, delivered by an electromagnet, was applied before and during polymerization. In these conditions, goethite nanorod orientation occurs before the crosslinking of the polymer matrix is completed. Series of samples at fixed goethite concentrations were prepared by applying different magnetic-field intensities (0; 0.18; 0.35; 0.7; 0.9; 1.5 T) during polymerization. The goethite volume fraction $\phi_H$ in the nanocomposite is used hereafter because it is a relevant physical parameter that affects orientational order in these materials. The values of goethite volume fractions for all samples are collected in Table 1.

**Characterization methods**.

The samples were examined with an Olympus BX51 polarizing microscope and their optical textures were photographed using an Olympus (Camedia C-3030) digital camera. Birefringence ($\Delta n$) measurements were performed under the microscope using a Berek (Olympus U-CBE) optical compensator. The sample birefringence, $\Delta n$, is an important physical property because, for single-domain samples, it is proportional to the nematic order parameter S: $\Delta n = \Delta n_{sat} \phi_H S$ where $\Delta n_{sat}$ is the intrinsic birefringence of the nanorods (See the Supplementary Information for more details on the determination of S from birefringence data). S is defined as $S = \frac{1}{2}(3\cos^2\varphi - 1)$ where $\varphi$ is the angle of a rod with the uniaxial symmetry axis **n** of the phase, defined by the magnetic field, and the brackets mean averaging over all rods. By definition, $S$ takes values ranging between $-1/2$ and 1, with $S = 0$ corresponding to the disordered isotropic phase. When the nematic order parameter is positive, $0 < S < 1$, the rods point on average along the axis **n**. Negative values of $S$ are the sign of so-called "antinematic" order[20] where the rods lie on average in the plane perpendicular to **n**, with isotropic angular in-plane distribution.

Small-angle X-ray scattering (SAXS) experiments were performed at the BM02 experimental station of the European Synchrotron Radiation facility at Grenoble, France. The sample-to-detection distance was 1.62 m and the wavelength was $\lambda = 0.116$ nm. The beam size was $0.4 \times .4$ mm². The data were recorded using a CCD Peltier-cooled camera (SCX90-1300, Princeton Instruments) of 1340×1300 pixels. Data processing (dark current subtraction, flat-field correction, grid distortion, and normalization) was performed using homemade software. The accessible range of scattering vector modulus q (q = $(4\pi\sin\theta)/\lambda$, where $2\theta$ is the scattering angle) was $0.056 < q < 1.13$ nm$^{-1}$. The nematic order parameter S can also be calculated from the anisotropy of the SAXS patterns by using a classical procedure (See the Supplementary Information for more details on the determination of S from X-ray scattering data) based on a convolution of the rod form factor with the orientational distribution function.[21]

Field Emission Gun Scanning Electron Microscopy (FEG-SEM, Hitachi SU-70, accelerating voltage 1 kV) was used to examine the organization of the nanoparticles in the composites. For this purpose, self-standing, ribbon-shaped, polymerized samples were carefully extracted from the flat glass capillaries. Images of the flat surface and the cross-section of these thin (50 μm thick) samples were recorded by setting either the sample surface or its section perpendicular to the electron beam.

### 3. Results and discussion

Homogeneous and transparent nanocomposites were obtained whatever the goethite volume fraction and the magnetic field applied during the polymerization. As previously described in detail,[11] the nanocomposites are comprised of nanoparticles embedded within a continuous polymer matrix. The presence of the iron oxide nanoparticles does not affect the conversion of HEMA in PHEMA (see Section 1 of the Supplementary Information) and the shape of the nanorods is not altered either. Moreover, the existence of strong interactions between the goethite nanoparticles and the side-groups of the PHEMA macromolecules have been previously evidenced by combining different analytical techniques (FTIR, ellipsometry) with equilibrium swelling experiments. It was thus demonstrated that a fraction of ester functions in the PHEMA chains are hydrolyzed and form carboxylate groups that interact with the surface of the nanoparticles. Such interactions ensure the good dispersion of goethite nanoparticles in the polymer matrix, regardless of their concentration and the magnetic fields applied. These specific interactions allow anchoring of the polymers to the fillers and avoiding macro-segregation and phase separation at high filler content, like in hybrid $SiO_2$-PHEMA matrix[22]. Consequently, a significant improvement of the mechanical properties of the nanocomposites has been observed[11].

Four samples, with $\phi_H$ = 1.08 or 4.31%, without magnetic field or with high magnetic field (1.5T) applied during polymerization, have been observed in FEG-SEM in order to characterize the nanorod organization in the polymer matrix at a sub-micronic scale. Images of the flat surface of the samples (Figure 1a,b), in contact with the flat faces of the glass capillary, reveal a random orientation of the nanorods, without any alignment even when polymerization was performed under high magnetic field. In contrast, alignment of the rods is clearly observed in the bulk of the sample, as illustrated by the image of the ribbon cross section (Figure 1c). Therefore, close to the surface of the sample, a continuous variation of the particle orientation, over a depth of about 100 nm, shows the influence of surface effects (Figure 1d). The competition between the rod alignment induced by the surface and that induced by the field in the bulk gives rise to this distorted region, at the surface, the thickness of which should decrease with increasing field intensity. The effect of the magnetic field on the orientation of the particles in the bulk is shown in Figure 2 for lower filler content ($\phi_H$ = 1.08%). While the rods are completely disoriented for composites polymerized without field, rod alignment is observed in high field even at such low volume fractions.

In order to probe particle alignment at a more macroscopic scale, we examined the optical textures of a goethite/PHEMA hybrid composite sample ($\phi_H$ = 4.31%, B = 0.18 T) observed by polarized-light microscopy (Figure 3). The texture is very uniform and much darker when the sample axis is parallel to either the polarizer or analyzer directions (Fig. 3a) than when it makes an angle of 45° with them (Fig. 3b). This optical anisotropy is the sign of a strong alignment of the goethite nanorods throughout the whole sample, at a millimetric length scale, even though the rod volume fraction is well below that of the nematic phase. Such an observation can be interpreted as follows: even in the isotropic phase, goethite nanorods partially align under the influence of the magnetic field,[18] in agreement with the FEG-SEM observations. This alignment was held by the magnetic field during polymerization of HEMA under UV-light. Finally, when the magnetic field was removed, the alignment of the goethite nanorods remained due to the glassy behavior of the PHEMA matrix at room temperature (Tg=100°C for the neat polymer).

Detailed analysis of the birefringence of the samples revealed that, upon HEMA polymerization, the nanorods were trapped with their main axis aligned parallel to the magnetic field at low field intensities and perpendicular to the field at high field intensities, as expected. Almost no birefringence was detected in the hybrid materials polymerized in a field of ~ 0.35 T. This behavior is indeed similar to that of the isotropic aqueous suspensions of goethite nanorods submitted to a magnetic field.[18]

The birefringence of all the samples was measured in order to investigate quantitatively the dependence of the nematic order parameter of the hybrid materials on the magnetic field intensity under which they were polymerized (Figure 4, left). We first observed that the samples polymerized in the absence of field nevertheless displayed a weak birefringence. This is most probably due to residual stresses arising from polymerization of the system in the confined environment of the flat capillary. Otherwise, the alignment of the nanorods in the composites, removed from the field, closely follows the field-induced alignment in the isotropic aqueous suspensions of nanorods (Figure 4a, inset). In particular, small (≈ 0.1-0.2) positive values of S are measured at low field intensities, S vanishes around 0.35 T, and large (≈ -0.4) negative values of S are measured at high fields. The nematic order parameter, S ≈ -0.4 at high fields is actually very large since the maximum value is -0.5 for this type of "antinematic" orientation.[20] This means that most nanorods are very well aligned in the plane perpendicular to the magnetic field direction.

These conclusions are confirmed by our SAXS measurements that probe nanorod alignment at a local scale but averaged over the illuminated sample volume (~ 1mm$^3$). The SAXS patterns (Figure 5) display a diffuse scattering halo that is most often anisotropic. The scattered intensity regularly decreases with scattering vector modulus, which suggests that the inter-particle interferences are negligible so that the scattering is mostly governed by the nanorod form factor. The scattering is stronger in the direction perpendicular to the magnetic-field direction at low field and in the parallel direction at high field. Again, this means that the nanorods are aligned parallel to the field below 0.35 T and perpendicular to the field beyond this value.

The nematic order parameter can classically be derived from the SAXS patterns through a fit of azimuthal circular scans (at constant scattering vector modulus) of the scattered intensity, assuming a Maier-Saupe orientational distribution of the nanorod axes.[21] The values of S of the hybrid materials are displayed in Figure 4 (right), as a function of the field intensity applied during polymerization. These values are quite consistent with those obtained from the birefringence measurements. The very good agreement between the SAXS and birefringence data proves that the latter method yields reliable values for the order parameter of the particles. This technique is much more easily implemented than SAXS or TEM and only requires knowledge of the optical properties of the particles (intrinsic birefringence) and of the matrix (refractive index). As such, it can be extended to a wide variety of composites based on anisotropic particles.

The persistence for months of the alignment of goethite nanorods at low volume fractions in the composite materials, after field removal, can be explained by the glassy behavior of the polymer matrix. Otherwise, the nanorod alignment may have vanished over some time. Moreover, since no influence of temperature was observed, up to ~ 100°C, above the glass transition of PHEMA, the nanorods must strongly interact with the polymer network, as previously inferred.[11, 22] This strongly suggests that the glass transition of PHEMA is higher around the goethite particles due to the surface specific interactions at the hybrid interface. These interactions lead to the formation of an interphase that improves the load transfer between the matrix and the fillers.[11] Therefore, the environment of the particles is too stiff to allow any motion.

Compared to more common epoxy/clay composites, goethite/PHEMA composites can be oriented with lower magnetic fields using simple and cheap commercial (NdFeB) permanent magnets that routinely provide about one Tesla. Thus, large areas may be aligned, which is more difficult to achieve with electromagnets. Moreover, epoxy/clay composites could also be aligned using electric fields[7] but this approach that involves direct contact with electrodes is much more invasive than the use of magnetic fields which do not require any contact with the samples. Furthermore, in contrast with clay/polymer nanocomposites aligned by shear flow, where the clay nanosheets are parallel to the flow, two different types of particle orientation, parallel or perpendicular to the field, could be achieved by exploiting the very peculiar magnetic properties of goethite nanorods.

Very importantly, in principle, the use of the isotropic phase rather than the nematic one has many advantages. There is no need to reach the concentration threshold beyond which the nematic phase spontaneously forms. In the case of nanoparticles of moderate aspect ratio such as the goethite nanorods considered here, the volume fractions required may be of the order of 10% whereas large alignment effects were already observed here at volume fractions of typically a few %, well below the nematic threshold. Moreover, the nematic phase always has a spontaneous nematic order parameter of $S \sim 0.75$-$0.85$ whereas, in principle, the order parameter induced by the field in the isotropic phase can be tuned at will, by increasing the field intensity, to any value $0 < S < 1$ for parallel alignment and $0 > S > -0.5$ for perpendicular alignment. In turn, all alignment-dependent properties (such as birefringence) can therefore be easily tuned. In addition, there is no need in this approach to address the recurrent problem of the control of the nematic texture, which requires expensive and delicate anchoring layers and surface treatments. Finally, spatial patterning of the samples can easily be achieved either by simply moving the sample in the field or by propagating a polymerization front while modulating the field intensity, as recently illustrated in the case of clay/polyacrylamide hydrogels submitted to a periodic electric field.[23]

Finally, we showed that the degree of particle alignment can be robustly determined using optical birefringence measurement. This latter technique is much easier to implement than the usual small-angle scattering method that is classically employed to characterize orientational order in nanocomposites, provided that the optical indices of the nanoparticles are known and the system is transparent.

ACKNOWLEDGMENT.

These experiments were performed on the BM02 beamline (CRG D2AM) at the European Synchrotron Radiation Facility (ESRF), Grenoble, France (beamtime request 2-01-799). This work was funded by the French ANR agency (Agence Nationale de la Recherche, Programme P2N, NASTAROD and Programme PNANO, MECHYBRID). Nicolas Chemin is kindly acknowledged by the authors for helpful discussions.

Table 1: Details of sample composition (contents of goethite and HEMA, corresponding goethite volume fraction $\phi_H$) (HEMA*: HEMA + 2 wt. % of DMPA)

| $\phi_H$ | Goethite suspension (mg) | HEMA* (mg) |
|---|---|---|
| 1.08 | 58 | 363 |
| 2.07 | 55 | 160 |
| 4.31 | 122 | 122 |
| 5.58 | 117 | 70 |

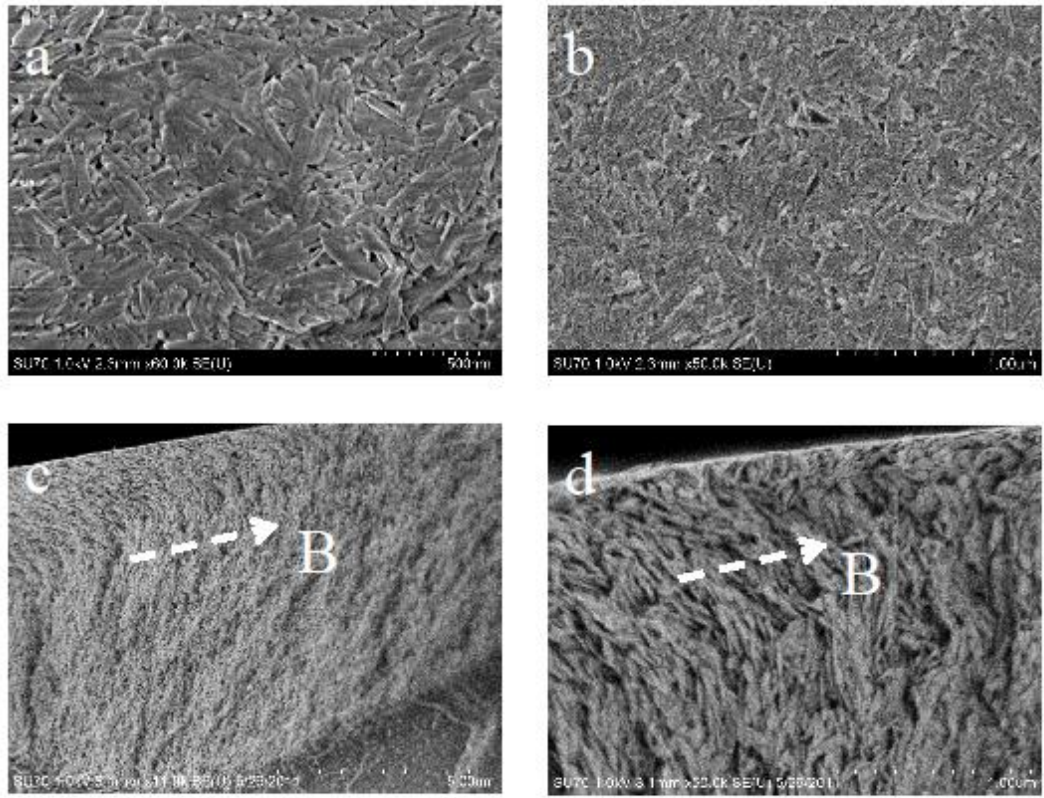

Figure 1: SEM-FEG images of the flat surface of nanocomposite samples with $\phi_H$=4.31%, polymerized in the absence of a magnetic field (a) and under a 1.5 T magnetic field perpendicular to the image (b) and of the cross section of the sample with $\phi_H$=4.31% polymerized under a 1.5 T magnetic field (dashed arrow), at two different magnifications (c) and (d).

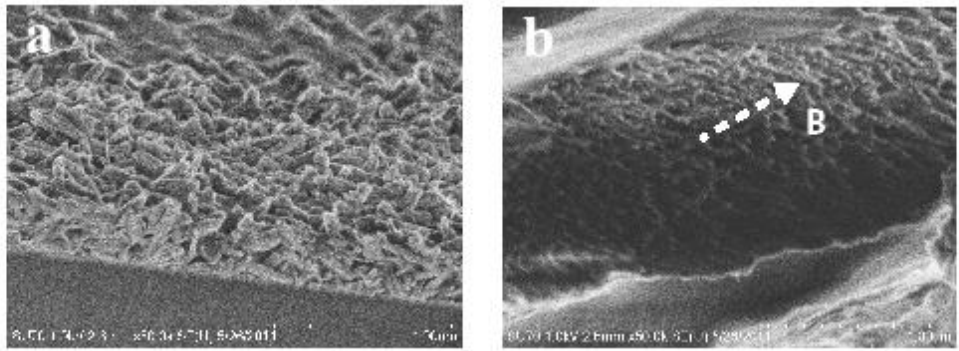

Figure 2: SEM-FEG images of the cross section of samples with $\phi_H$ = 1.08% samples polymerized (a) in the absence of a magnetic field and (b) under a 1.5 T magnetic field (dashed white arrow).

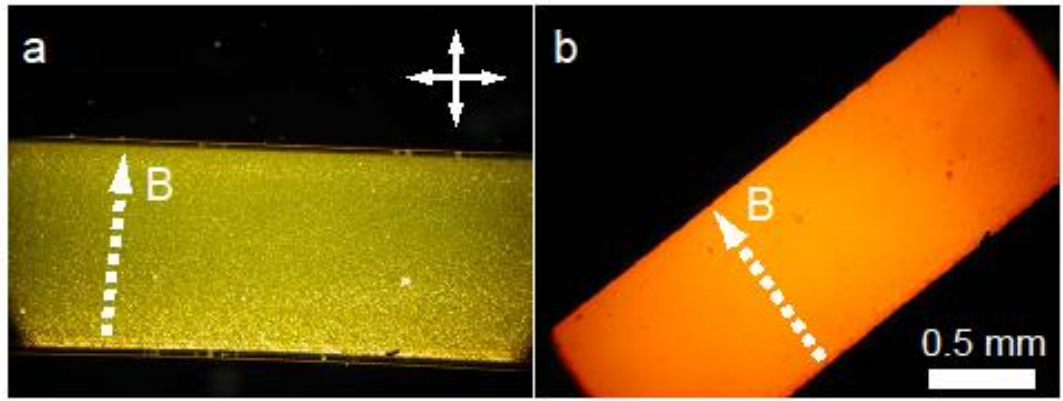

Figure 3: Images in polarized-light microscopy of a sample with $\phi_H$ = 4.31% samples polymerized under a 0.18 T magnetic field (white dashed line) perpendicular to the flat glass capillary axis. (No magnetic field was applied during the microscopic observations.). The capillary appears rather dark when its main axis is parallel the directions either of the polarizer or of the analyzer (white double arrows) and it appears quite bright when the capillary axis makes an angle of 45° with the directions of the polarizer and the analyzer, which illustrates the good alignment of the sample.

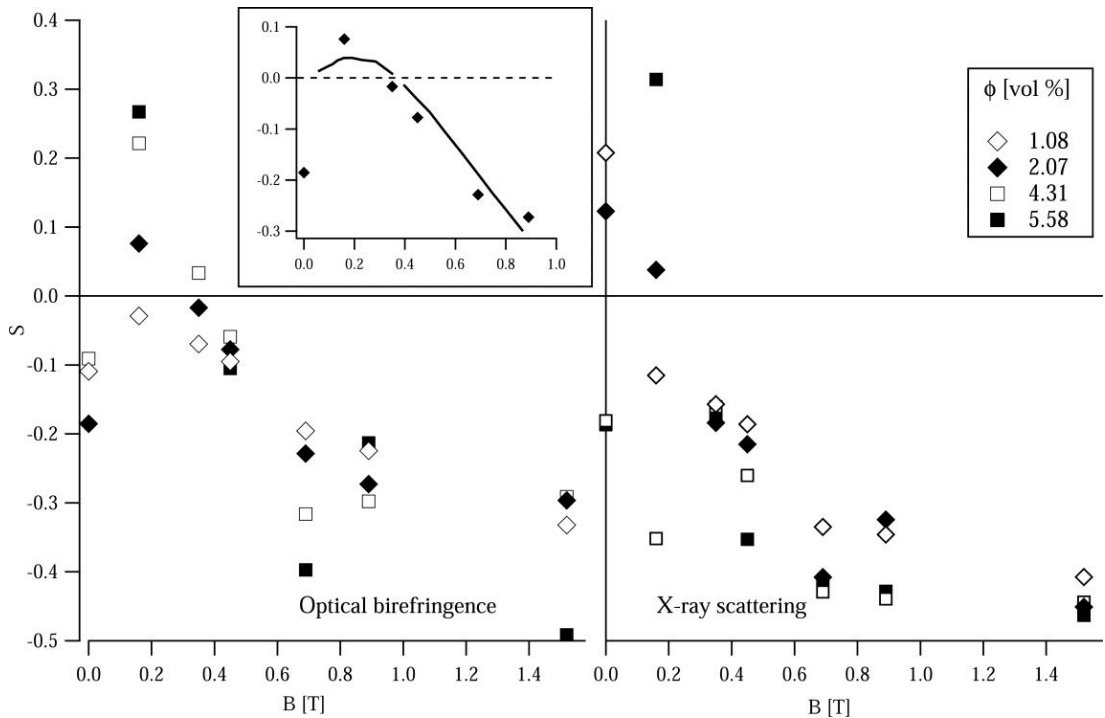

Figure 4: Order parameter S as a function of the magnetic field B applied during polymerization. (Left). Determined from the optical birefringence of the sample, $\Delta n$. The inset shows a comparison, for $\phi_H$ = 2.07 %, of the orientational order (diamonds) of the nanorods in the composites, removed from the field, with the field-induced orientational order (solid line) in the aqueous suspensions of goethite nanorods, at $\phi_H$ = 2.10 % . (Right). Determined from the SAXS patterns. Different symbols correspond to different volume fractions of goethite, $\phi_H$, indicated in the legend.

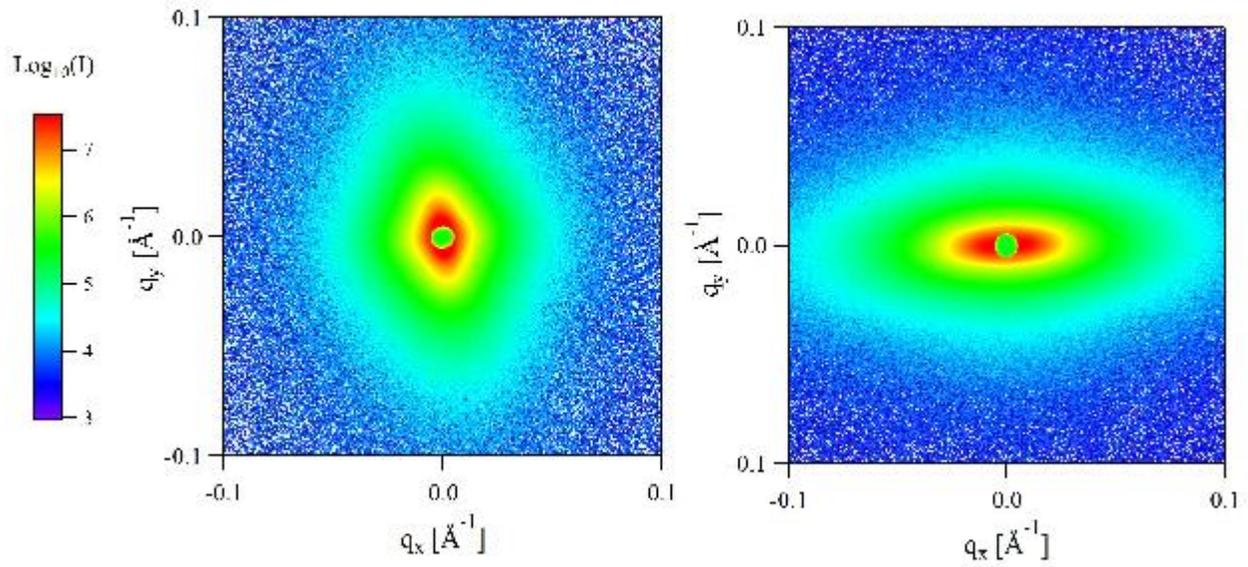

Figure 5: SAXS patterns for two samples polymerized under magnetic field (applied along the horizontal direction). (No magnetic field was applied during the X-ray scattering measurements.) Left: $\phi_H$ =5.58 % and $B$ = 0.2 T (alignment along the field, $S > 0$). Right: $\phi_H$ =4.31 % and $B$ = 1.5 T (alignment perpendicular to the field, $S < 0$).